\documentclass [12pt]{article}
\usepackage {graphicx}
\usepackage {longtable, cite}
\begin{document}


\begin{center}
\textbf{ON INFLUENCE OF INTENSIVE STATIONARY ELECTROMAGNETIC FIELD ON THE
BEHAVIOR OF FERMIONIC SYSTEMS}
\end{center}

\begin{center}
V.N. Rodionov, A.M. Mandel, E.V. Arbuzova \\ \textit{General
Physics
Department, \\ Moscow State Geological Prospecting Academy, \\
117997, Miklukho -Maklaya, 23}
\end{center}

\begin{center}
\textbf{Abstract.}
\end{center}

\textit{Exact solutions of Schr\"{o}dinger and Pauli equations for
charged particles in an external stationary electromagnetic field
of an arbitrary configuration are constructed. Green functions of
scalar and spinor particles are calculated in this field. The
corresponding equations for complex energy of particles bounded by
short range potential are deduced. Boundary condition typical for
$\delta$ - potential is not used in the treatment. Explicit
analytical expressions are given for the shift and width of a
quasistationary level for different configurations of the external
field. The critical value of electric field in which the idea of
quasistationary level becomes meaningless is calculated. It is
shown that the common view on the stabilizing role of magnetic
field concerns only scalar particles.}

\begin{center}
\textbf{I. Introduction. Formulation of the problem.}
\end{center}

The influence of external electromagnetic fields on
nonrelativistic reactions with charged particles and on the
behavior of bound systems (like atoms, ions and atomic nucleuses)
has being investigated systematically for a long time see, (for
example, ~\cite {Kel.1, Fr, Zel, Dem.1, Dem.2, Kel.2, Ser, Asp,
Rit, Druk, Nik, Baz, Zeld, Pop, Man} ). However, a set of
problems, in our opinion, requires additional studying. First, it
is the appearance of physically meaningless divergences in
equations for the complex energy of quasistationary bound states.
Second, the question about the behavior of fermionic systems in
super strong electromagnetic fields is not enough elucidated. Note
that the basic results in this field of physics were received by
nonperturbative methods. Usually the exact solutions of
Schr\"{o}dinger equations with Hamiltonians taking into account
the presence of external fields are used ~\cite {Dod, Kl}. It is
important, that the field can become intensive not only due to
increasing of the strength, but also due to decreasing of binding
energy of the system. In other words, in the vicinity of the
reaction threshold an arbitrary external field is strong. Third,
the common view on the stabilizing role of magnetic field in
decays of quasistationary states is inadequate for spinor
particles ~\cite {RoKr1, RoKr2}. That view is caused by the fact
that spinor states of electrons usually are not taken into
accounts in nonrelativistic reactions. In this paper we treat a
part of these deficiencies.

Let us consider a charged particle bounded by a short range
potential like a $\delta$-potential and located in an external
stationary electromagnetic field of an arbitrary configuration.
Note that the potential of zero-radius is a rather widely spread
in the literature approximation for a multi-electronic atom field
and especially for a negative ion field, and also for the field of
nuclear forces ~\cite {Baz, Dem.3}. In the general case the
external field is defined by three independent parameters:
strengths of the magnetic \textit{H} and electric \textit{E}
fields and the angle \textit{ $\phi$} between them.

The process of particle coming out of a $\delta$-well leads to the
transition from an isolated energy level to a level in a
continuous spectrum, \textit{i.e.} to a decay of the bound state.
The external electromagnetic field influences on this transition
in two ways. First, it forms wave functions of the ``free'' state
and, second, shifts and broads the bound level in a $\delta$-well.
As the result the initial state of the particle in the external
field becomes quasistationary. The most adequate instrument for
the investigation of such states is the complex energy formalism
(for example, ~\cite {Dem.1, Baz, Lan})

\begin{equation}
\label{eq1} W = W_{0} + \Delta W - i\Gamma/2 ,
\end{equation}
where $W_{0}<0$ is the energy of the nonperturbative bound level,
 $\Delta W$ is the real shift caused by the external field, and $\Gamma$ is
the width of a level associated with the decay probability of the
bound state. It is important that a separate consideration of the
level shift and its width is possible only in a weak external
field. In an intensive field it is necessary to investigate these
quantities in common.

The main purpose of this article is to derive an equation for the
complex energy in a stationary external field by the
\textit{method of an analytical continuation} and to solve it for
some field configurations. This treatment is principally different
from the traditional derivation using the boundary condition
typical for the $\delta$-well ~\cite {Dem.1, Pop, Man}.

The general structure of the paper and its main results are as
follows. In the second Section explicit solutions for
Schr\"{o}dinger and Pauli equations are constructed both for
scalar and spinor particles in an external stationary
electromagnetic field. Then Green functions of the particles are
constructed with its help. In spite of rather complicated
configuration of the external field, these functions can be
represented in terms of elementary functions. That substantially
simplifies their further applications.

In the third Section the expression for the bound level width in
the external field is deduced with help of the Green functions.
Then the equation for the complex energy of the bound particle in
the $\delta$-well is built by the method of analytical
continuation. It has several features different from the ones of
the equation traditionally used. First, our equation is not
transcendental and it is explicitly resolved with respect to the
complex energy. That is why its right handed side (one fold
integral) does not contain exponential divergences typical for
equations used earlier. Thus it does not require any additional
regularization. Second, we consistently take into account the spin
states, which is fundamentally important in the presence of
magnetic field.

Finally in the fourth Section the equation for the complex energy
is solved in several cases of physical interest. It is well known
that the electric field destroys the bound state of a charged
particle in a potential well. Thus we calculated the critical
value of the strength of the electric field in which the width of
the level becomes comparable with its depth in the potential well.
In such a field the notion of an isolated level loses meaning
~\cite {Lan}. When the strengths of the electric and magnetic
fields are comparable the latter one gives only small corrections
to the usual effects: the Stark shift and the broadening due to
the tunnel effect. But the signs of these corrections are
different for scalar and spinor particles. In the case when the
magnetic field is dominating the behaviors of the complex energy
of the mentioned particles are principally different. For scalar
particles the magnetic field exponentially decreases the bound
level width and therefore stabilizes it. It was assumed that the
same effect takes place also for electrons (e.g. ~\cite {Dem.2,
Druk, Pop, Man}). But as was shown by the authors of this paper in
~\cite {RoKr1, RoKr2}, the strong magnetic field linearly
increases the level width for particles with spin $1/2$. Hence in
the presence of an external magnetic field it is necessary to take
into account the spin of the given particle and its contribution
to the complex energy of the quasistationary level even in
non-relativistic problems.

\begin{center}
\textbf{II. The wave function and the Green function of a charged
particle in an external stationary electromagnetic field}
\end{center}

In order to receive the wave function of a free charged spinless
particle in a stationary electromagnetic field of a general
configuration, it is necessary to solve the Schr\"{o}dinger
equation with the following Hamiltonian:

\begin{equation}
\mathord{\buildrel{\lower3pt\hbox{$\scriptscriptstyle\frown$}}\over
{H}} _{Sh}  = - \frac{{\hbar ^{2}}}{{2m}}\Delta - i\hbar \omega
_{H} x\frac{{\partial} }{{\partial y}} + \frac{{1}}{{2}}m\omega
_{H} ^{2}x^{2} + eEx\sin\varphi  + eEz\cos\varphi ,
\end{equation}
where \textit{e} and \textit{m} are the charge and the mass of the
particle; \textit{E} and \textit{H} are the strengths of the
electric and the magnetic fields, respectively; $\varphi$ is the
angle between the corresponding vectors;  $\omega_{H} = eH/mc$ is
the cyclotron frequency. We suppose that the magnetic field has
the orientation along \textit{z} axis, and the vector of the
electric field strength lies in the \textit{xz} plane.

The Schr\"{o}dinger equation with the Hamiltonian (2) can be
solved by the standard method of variables separation. The moving
of the particle in a plane normal to the magnetic field has the
same character as the one in the absence of electric field.
Corresponding eigenfunctions are Hermit functions, each of them is
related to a certain Landau level. In this case the coordinate
$x_{0}$ of the orbit centre depends on the conserved transverse
momentum  $p_{y}$ and on the transverse component of the electric
field $E cos\varphi$. The movement of the particle along
\textit{z} axis is similar to the one in the uniform electric
field, and it can be described by Airy functions ~\cite {Lan}. It
is not difficult to see that the wave function of the charged
particle in the field of the considered configuration has the form

\begin{equation}
\psi _{n} \left( {\vec {r},t} \right)\, = \,N\,u_{n} \left( {\rho}
\right)\,{\rm B}\left( {\xi}  \right)\exp\left[
{\frac{{i}}{{\hbar} }\left( {yp_{y} - t\tilde {W}} \right)}
\right] ,
\end{equation}
where \textit{N} is a normalization coefficient, the Hermit
function $u_{n} \left( {\rho}  \right)$ satisfies the equation

\[
{u}''_{n} ({\rho}) + ( {2n + 1 - \rho} )u_{n} ({\rho}) = 0
\]
where \textit{n} is the number of the Landau level, and the
argument of this function is

\[
\rho \left( {x} \right)\,\, = \left( {\frac{{m\omega _{H}}
}{{\hbar} }} \right)^{1/2}\left[ {x + \frac{{p_{y}} }{{m\omega
_{H}} } - \frac{{eE\sin\varphi} }{{m\omega _{H} ^{2}}}} \right] .
\]
Function $B \left( {\xi} \right)$ in (3) is proportional to the
Airy function \textit{Ai} regular at infinity

\[
{\rm B}\left( {\xi}  \right)\,\, = \,\frac{{\left( {2m}
\right)^{1/3}}}{{\pi \left( {eE\cos\varphi} \right)^{1/6}\hbar
^{2/3}}}Ai\left( {\xi } \right)
\]
with the argument

\[
\xi (z) =  \left( {\frac{2meE\cos\varphi}{\hbar ^{2}}}
\right)^{1/3} (z - z_{0}) ,
\]
where $z_{0}$ is defined by the total energy \textit{W} of the
particle in the external field

\begin{equation}
\tilde {W} = \hbar \omega _{H} \left( n + \frac{1}{2} \right) + eE
z_{0} \cos\varphi .
\end{equation}

The solution of the Schr\"{o}dinger equation constructed in that
way is valid for a scalar particle. In order to take into account
spin states of the particle in the concerned non-relativistic
approximation it is necessary to pass on to the Pauli equation
(see, for example, ~\cite {Lan}) for the spinor wave function
$\bar {\psi} _{n \sigma}  \left( {\vec {r},t} \right)$

\[
i\hbar \frac{\partial}{\partial t} \bar {\psi}_{n \sigma} \left(
{\vec {r},t} \right) =
\mathord{\buildrel{\lower3pt\hbox{$\scriptscriptstyle\frown$}}\over
{H}} _{P\,} \bar {\psi}_{n \sigma}  \left( {\vec {r},t} \right) ,
\]
where the Hamiltonian of the Pauli equation involves the energy of
the spin interaction with the external stationary magnetic field
$\vec {H}$:

\[
\mathord{\buildrel{\lower3pt\hbox{$\scriptscriptstyle\frown$}}\over
{H}} _{P\,} =
\mathord{\buildrel{\lower3pt\hbox{$\scriptscriptstyle\frown$}}\over
{H}} _{Sh} + \frac{{e\hbar} }{{2mc}}\vec {\sigma} \vec {H} ,
\]
and $\vec {\sigma}$ are the Pauli matrixes. Taking into account
the concrete orientation of the magnetic field this Hamiltonian is
reduced to the matrix form

\[
\mathord{\buildrel{\lower3pt\hbox{$\scriptscriptstyle\frown$}}\over {H}}
_{P\,} = \left( {{\begin{array}{*{20}c}
 {\mathord{\buildrel{\lower3pt\hbox{$\scriptscriptstyle\frown$}}\over {H}}
_{Sh} + \hbar \omega _{H} /2} \hfill & {0} \hfill \\
 {0} \hfill &
{\mathord{\buildrel{\lower3pt\hbox{$\scriptscriptstyle\frown$}}\over {H}}
_{Sh} - \hbar \omega _{H} /2} \hfill \\
\end{array}} } \right).
\]

If we put now into the Pauli equation the spinor wave function of the mixed
state with an arbitrary time dependence of the coefficients

\begin{equation}
\bar {\psi}_{n \sigma} \left( {\vec {r},t} \right) = \psi_{n}
\left( {\vec {r},t} \right) \left( {{\begin{array}{*{20}c}
 {a_{1} \left( {t} \right)\exp\left( { - i\omega _{H} t/2} \right)} \hfill \\
 {a_{2} \left( {t} \right)\exp\left( {i\omega _{H} t/2} \right)} \hfill \\
\end{array}} } \right) ,
\end{equation}
where the scalar function $\psi_{n} \left( {\vec {r},t} \right)$
is defined by the formula (3), we get

\[
da_{1}/dt = da_{2}/dt = 0 .
\]

It means, first, that the spin projection on the \textit{z} axis
is conserved. This conservation is provided by the homogeneity and
the stationarity of the magnetic field. Second, the energy of the
particle is defined, instead of the expression (4), by the more
common formula

\begin{equation}
\tilde {W} = \hbar \omega _{H} (n + 1/2 + \sigma) + eE
z_{0}\cos\varphi ,
\end{equation}
where $\sigma = 0$ for the scalar particle and $\sigma = \pm 1/2$
for the spinor one.

Thus four constants ($p_{y}, z_{0}, n, \sigma$) unambiguously
parameterize the wave function of the particle with spin 1/2 in
the external field under consideration.

Let's construct now a time depending Green function of a charged particle in
the given field. With the purpose of some simplification of calculations we,
without limiting the generality, suppose one of arguments of point-to-point
Green function to be zero. The function will be a diagonal matrix of a kind

\begin{equation}
\mathord{\buildrel{\lower3pt\hbox{$\scriptscriptstyle\frown$}}\over
{G}} \left( {\vec {r},t;\,\vec {0},0} \right)\,\, =
\,\sum\limits_{n = 0}^{\infty } {\int\limits_{ - \infty} ^{\infty}
{dz_{0}}  \int\limits_{ - \infty }^{\infty}  {dp_{y} G_{Sh}
\,\left( {n,\,z_{0} ,\,p_{y}}  \right)\left(
{{\begin{array}{*{20}c}
 {\exp\left( { - i\omega _{H} t/2} \right)\quad \quad 0} \hfill \\
 {0\quad \quad \quad \exp\left( {i\omega _{H} t/2} \right)} \hfill \\
\end{array}} } \right)}}  \
\end{equation}
where the scalar Green function $G_{Sh} \left( {n,z_{0} ,p_{y}}
\right)$ corresponds to a Schr\"{o}dinger equation

\[
G_{Sh}\left( {n, z_{0}, p_{y}} \right) \sim u_{n}(x) u_{n}(0)
Ai(z) Ai (0) \exp\left[ \frac{i}{\hbar} \left( p_{y} y - t\tilde
{W} \right) \right].
\]

The integral over $z_{0}$ included in (7) is easily calculated
using the integral representation for the Airy function. The
integral over $p_{y} $ with Hermit functions is similar to the one
that was considered in Ref. ~\cite {Klep}. The series in Lagerre
polynomials formed after that is summarized over \textit{n} with
help of the algorithm described in ~\cite {Rod1} (see also ~\cite
{Fey}). As a result the scalar part of the Green function in a
stationary electromagnetic field is expressed in terms of
elementary functions:

\begin{equation}
 G_{Sh} \left( {\vec {r},t;\,\vec {0},0} \right) = \left( {\frac{{m}}{{2\pi
 i}}} \right)^{3/2}\frac{{\omega _{H}}
}{{2t^{1/2}}}\sin^{ - 1}\,\left( {\frac{{\omega _{H} t}}{{2}}}
\right)\exp\left( {\frac{{iS}}{{\hbar} }} \right),
\end{equation}
where

\begin{eqnarray}
 S = \frac{mz^{2}}{2t} - \frac{eEzt}{2} \cos\varphi
 - \frac{(eE\cos\varphi)^{2}t^{3}}{24m} +
\frac{m\omega_{H}}{4} \left[ (x^{2} + y^{2})\cot
\left(\frac{\omega_{H} t}{2}\right )
 - 2xy \right]\nonumber\\
 + \frac{1}{2}eExt\sin\varphi  + \frac{eE\sin\varphi}
 {\omega _{H}}\left[\frac{\omega _{H} t}{2} \cot \left( \frac{\omega
_{H} t}{2}\right) - 1 \right] \left( y +
\frac{eEt\sin\varphi}{2m\omega _{H}} \right). \phantom{1-1}
\end{eqnarray}

If we assign in (8), (9) the mass of the particle to unity and
perform obvious changes in notation we exactly reproduce formulas
(A.7) - (A.9) from Ref. ~\cite {Pop}. At $\varphi = {{\pi}
\mathord{\left/ {\vphantom {{\pi}  {2}}} \right.
\kern-\nulldelimiterspace} {2}}$ the relations (8), (9) describe
the Green function of a scalar particle in the crossed field
~\cite {Rit, Druk, Fey}.

\begin{center}
\textbf{3. Derivation of the equation for the complex energy of a bound
particle in a stationary external field by the method of analytical
continuation}
\end{center}

The expression for the probability \textit{P} of the decay of a
particle bound state per unit time (or for the width $\Gamma $ of
the bound level) in the considered field can be received with help
of the Green function, using the procedure developed in Refs.
~\cite {Rod2, Nik}. This probability has the form

 \begin{eqnarray}
 P = \pi \left(\frac{2}{m}\right)^{3/2} \left(\frac{\left|W_{0}
\right|}{\hbar}\right)^{1/2}\int\limits_{ - \infty} ^{\infty} {dt}
G_{Sh} (\vec 0,t; \vec 0,0) f_{\sigma}\left( \frac{\omega
_{H}t}{2}\right) \exp \left( \frac{i}{\hbar} W_{0}  t \right) \nonumber\\
 = \frac{\omega _{H}}{2}\left( \frac{\left| W_{0} \right|}{\pi
 \hbar} \right)^{1/2} \exp\left( { - i\frac{{3\pi}
}{{4}}} \right)\int\limits_{ - \infty} ^{\infty}
{\frac{{dt}}{{t^{1/2}}}} \frac{{f_{\sigma}  \left( {\omega _{H}
t/2} \right)}}{{\sin\left( {\omega _{H} t/2} \right)}}\exp\left[
{\frac{{i}}{{\hbar} }\left( {S + W_{0}  t} \right)} \right],
\end{eqnarray}
where $W_{0}$ is the energy of the unperturbed bound level
(\ref{eq1}), and \textit{S} at zero coordinates is defined by the
formula

\begin{equation}
S = - \frac{{\left( {eE\cos\varphi}  \right)^{2}t^{3}}}{{24m}} +
\frac{{\left( {eE\sin\varphi}  \right)^{2}t}}{{2m\omega _{H}
^{2}}}\left[ \frac{\omega _{H} t}{2}\cot\left(\frac{\omega _{H}
t}{2}\right) - 1 \right].
\end{equation}

Poles of the integrand function in points $t_{n} = {{2\pi n}
\mathord{\left/ {\vphantom {{2\pi n} {\omega _{H}} }} \right.
\kern-\nulldelimiterspace} {\omega _{H}} }$ are passed from below.

Function $f_{\sigma} (a)$ is identically equal to unity for scalar
particles, but for spinors it depends on the polarization of the
bound particles in the initial state and of free particles in the
final one. In the general case

\begin{equation}
f_{\sigma} (\omega _{H} t/2) = a_{1}^{2} \exp ( - i\omega _{H} t/2
) + a_{2}^{2} \exp ( i\omega _{H} t/2 ),
\end{equation}
where $a_{1} (a_{2})$ are the amplitudes of the probability
introduced in (5), that the spin of a particle is directed along
(or against) the magnetic field. The normalization condition
demands that $a_{1}^{2} + a_{2}^{2} = 1$. For example, the most
realistic situation in the process of ionization is the one where
bound electrons are not polarized and the detector does not
distinguish between polarizations of free electrons. In this case
for the calculation of the total probability of ionization it is
necessary to average over polarizations of the initial particles
and to sum over polarizations of the final ones (see, for example,
~\cite {Lan}). Then $a_{1}^{2} = a_{2}^{2} = 1/2$ and
$f_{\sigma}=\cos(\omega _{H} t/2)$.

As it is known the width of the bound level is proportional to the
probability of its decay $\Gamma = \hbar P$. For analytical
continuation let us rewrite (10) in such a way that the
integration in it is carried out on the positive semi-axis. With
this purpose it is necessary to resolve the exponent in (10) into
the real and the imaginary parts and in the integral contained
$\cos\left[ {\left(S + W_{0} t\right)/\hbar}\right ]$ to bypass
the zero point from below on an infinitesimal contour. Such a
bypass cancels the divergence proportional to $t^{- 1/2}$ of the
above mentioned integral at zero. As a result the formula for the
level width takes the form

\begin{eqnarray}
\frac{\Gamma}{4 \left| W_{0} \right|^{1/2}} = \frac{1}{2}\left(
\frac{\hbar}{\pi} \right)^{1/2} \int\limits_{0}^{\infty}
{\frac{dt}{t^{3/2}}}\cdot \hphantom{1-1-1-1-1}\nonumber\\
\left\{ \frac{\omega _{H} t}{2\sin( \omega _{H} t/2 )}
\sum\limits_{k=1,2} a_{k}^{2} \sin\left[ \frac{1}{\hbar} \left(S +
W_{0}t + (-1)^{k}  \frac{\hbar \omega _{H}}{2} t \right) -
\frac{\pi}{4} \right]\ + \sin\frac{\pi}{4} \right\}.
\end{eqnarray}

It becomes obvious now that the right hand side of (14) represents
the imaginary part of the complex expression

\begin{eqnarray}
(- W)^{1/2} = \frac{1}{2} \left(\frac{\hbar}{\pi}\right)^{1/2}
\int\limits_{0}^{\infty} {\frac{dt}{t^{3/2}}}\cdot
\hphantom{1-1-1-1-1-1-} \nonumber\\
\left\{ \frac{\omega _{H} t}{2\sin (\omega _{H} t/2)}
\sum\limits_{k=1,2} a_{k}^{2} \exp\left[ \frac{i}{\hbar}\left( S +
W_{0} t + (-1)^{k} \frac{\hbar \omega _{H}}{2} t \right) -
\frac{i\pi}{4} \right] - \exp\left( - \frac{i\pi}{4} \right)
\right\}
\end{eqnarray}
and the left hand side is nothing else than the imaginary part of
the expansion (1)

\[
\left( { - W} \right)^{1/2} = \left( { - W_{0} - \Delta W +
i\Gamma /2} \right)^{1/2} \approx \left| {W_{0}}  \right|^{1/2} -
\frac{{\Delta W}}{{2\left| {W_{0}}  \right|^{1/2}}} +
i\frac{{\Gamma} }{{4\left| {W_{0}} \right|^{1/2}}}.
\]

It is logical to assume that the real part of (14) determines the
level shift due to the external field. If we use now the identical
transformation (recalling that $W_{0} < 0$)

\begin{equation}
\int\limits_{0}^{\infty} {\frac{dt}{t^{3/2}}} \left[ \exp\left(
i\frac{tW_{0}}{\hbar} \right) - 1 \right] = - 2\left( \frac{\pi
|W_{0}|}{\hbar} \right)^{1/2} \exp\left(\frac{i\pi }{4}\right),
\end{equation}
we receive the final equation for the complex energy of a bound
level in the stationary external field of a general configuration

\begin{eqnarray}
( - W)^{1/2} - (- W_{0})^{1/2}
 = \frac{1}{2}\left(\frac{\hbar}{i\pi}\right)
^{1/2}\int\limits_{0}^{\infty } {\frac{dt}{t^{3/2}}}
\exp\left(\frac{i}{\hbar} W_{0}t \right) \cdot \nonumber\\
\left\{\frac{\omega _{H} t}{2} \frac{f_{\sigma}(\omega_{H}
t/2)}{\sin(\omega_{H}  t/2)} \exp\left( \frac{iS}{\hbar }\right) -
1  \right\},\phantom{1-1-1}
\end{eqnarray}
where $S$ is defined by formula (11).

We see that the received equation is resolved explicitly with
respect to $W$. Note, that all integrals in intermediate
expressions used in the derivation as well as in the final formula
(16) \textit{are finite.} Besides the condition of the relative
smallness of the initial level width and shift has been used. It
imposes certain restrictions on the area of applicability of (16).
However, such restrictions are incorporated in the concept of a
quasistationary energy level ~\cite {Lan}. More strict
mathematical derivation of the equation similar to (16) is given
in ~\cite {Kad}.

In appendix A of paper ~\cite {Pop} a more commonly used variant
of derivation of the equation for the complex energy in the
considered configuration of an external field (see also ~\cite
{Dem.1, Dem.2, Druk, Man, Dem.3}) is described. Taking into
account the spins of particles, we can get the equation

\begin{eqnarray}
( - W)^{1/2} - (- W_{0})^{1/2}
 = \frac{1}{2}\left(\frac{\hbar}{i\pi}\right)
^{1/2}\int\limits_{0}^{\infty } {\frac{dt}{t^{3/2}}}
\exp\left(\frac{i}{\hbar} W t \right) \cdot \nonumber\\
\left\{\frac{\omega _{H} t}{2} \frac{f_{\sigma}(\omega_{H}
t/2)}{\sin(\omega_{H}  t/2)} \exp\left( \frac{iS}{\hbar }\right) -
1  \right\},\phantom{1-1-1}
\end{eqnarray}
which differs from the equation (16) only by the substitution
$W_{0} \to W$ in its right hand side. Thus the two approaches
leading to slightly different equations are formulated for the
description of quasistationary systems bounded by short-range
forces. In this connection it is useful to compare the areas of
their applicability and to understand how much are different their
solutions.

First note that usually an iteration method is used to solve
equation (17). Therefore the substitution $W_{0} \to W$ is
accepted on the first step of the iterative procedure, reducing
thereby (17) to (16). Therefore the received equation (16) is
treated usually as \textit{an approximate one}, working only in
the case of weak external fields. Thus, the choice of one of the
alternative approaches suggested above is practically reduced to
the question: whether iterative corrections of higher orders are
necessary or it is needed to stop on the ``first iteration''.

Obviously, the equation (17) for the complex energy is
transcendental and, hence, more complicated than (16). Solving a
more complicated equation is meaningful only if it gives a more
exact or an essentially new result. Therefore it is important to
discuss approximations used by the derivation of the specified
relations. At the derivation of equation (16) we used conditions
$\Gamma \ll |W_{0}|,  |\Delta W| \ll |W_{0}|$, which impose
certain restrictions from above on values of external fields.
However, the presence of such restrictions follows both from the
non-relativistic character of the problem, and from the initially
assumed presence of "the shallow bounded level'' in a deep narrow
potential well ~\cite {Baz}. Moreover, the concept of a
quasistationary level assumes the smallness of its decay
probability ~\cite {Lan}. We shall discuss the last circumstance
in detail below.

However, it is not possible at all to estimate restrictions on the
values of external fields, admitted at the derivation of (17). The
matter is that all intermediate integrals (formulas (A.6), (A.13)
- (A.19) in Ref. ~\cite {Pop}) as well as the final integral in
Eq. (17) \textit{are exponentially divergent} . At that it is
emphasized in works ~\cite {Dem.1, Dem.2, Dem.3} that the typical
for a $\delta$-well boundary condition is valid only for a complex
value of $W$

\begin{eqnarray}
G\left( {\vec {r},\vec {0};{\kern 1pt} {\kern 1pt} W}
\right)\mathop { \approx }\limits_{r \to 0} \frac{{m\hbar
^{1/2}}}{{2\pi \,r}} - \frac{{m^{3/2}}}{{\left( {2\hbar }
\right)^{1/2}\pi }}\left( { -
W} \right)^{ - 1/2} +  \hphantom{1-1-1-1}\\
i^{ - 1/2}\left( {\frac{{m}}{{2\pi }}}
 \right)^{3/2}\int\limits_{0}^{\infty
} {\frac{{dt}}{{t^{3/2}}} \exp\left( {i\frac{{Wt}}{{\hbar }}}
\right)} \left\{ {\frac{{\omega _{H} t}}{{2}} \sin^{ - 1}\left(
{\frac{{\omega _{H{\kern 1pt} } t}}{{2}}} \right) \exp\left(
{\frac{{i S}}{{\hbar }}} \right) - 1} \right\} + O\left( {r}
\right)\nonumber .
\end{eqnarray}

Therefore it remains not clear to terms of which order it is
necessary to relate the last integral at $Im W < 0$.

Finally, in our opinion, the key condition

\begin{equation}
G\left( {\vec {r},\vec {0};{\kern 1pt} {\kern 1pt} W}
\right)\mathop { \approx }\limits_{r \to 0} G_{0} \left( {\vec
{r},\vec {0};{\kern 1pt} {\kern 1pt} W_{0} } \right)\mathop {
\approx }\limits_{r \to 0} \frac{{m\hbar ^{1/2}}}{{2\pi \,r}} -
\frac{{m^{3/2}}}{{\left( {2\hbar } \right)^{1/2}\pi }}\left( { -
W_{0} } \right)^{ - 1/2} + O\left( {r} \right).
\end{equation}
on which total procedure of a derivation ~\cite {Pop} is based
causes some doubt. Namely the comparison of asymptotics (18) and
(19) gives the traditional equation for a complex energy (17). The
condition (19) actually demands that not only the singular part
but also the finite one of the asymptotic expansion of the Green
function at zero does not depend on external fields. As far as we
know, arguments of physical character for the benefit of this were
not given in the literature. Nevertheless the asymptotic of the
field Green function at zero (18) contains in the \textit{finite}
part the integral, divergent for complex values of $W$. In these
conditions the requirement of the equality of finite terms is an
obvious excess of the accuracy of the used approximation. This
will be shown in the following section for the case of a pure
electric external field.

The mentioned above difficulties of the derivation of expression
(17) are characteristic not only for an approach suggested in Ref.
~\cite {Pop}. They are typical for any derivation of equation (18)
using the boundary condition of the $\delta$-potential with a
complex energy. There are no such problems in Refs. ~\cite {Dem.1,
Dem.2, Druk} for the reason that their authors determine the
complex energy, in contrast to (\ref{eq1}), with a positive
imaginary part. It removes the problem of divergences in formulas
like (17)-(19), but contradicts to the classical definition of the
complex energy. In agreement with such a definition all
probabilities in a quasistationary state should exponentially
increase in time ~\cite {Lan}.

Thus, it does not seem to be possible, in our opinion, to derive
equation (17) with the complex energy $W$ in the right hand side
consistently and accurately from the mathematical point of view.
It is especially problematic to trace all approximations used in
such a derivation.

\begin{center}
\textbf{4. Solving equations for the complex energy of a bound
particle for some configurations of external fields.}
\end{center}

Let us stop now on some variants of the solution of the received
equations for simple configurations of an external field. It is
stated in Ref. ~\cite {Man1} that the elimination of divergences
in integrals like (18) by the substitution $W \to W_{0}$ is
incorrect. Let's discuss further a rather simple configuration of
an external electrostatic field from the point of view of the
validity and the applicability of equations (16) and (17). We use
the notation of Ref. ~\cite {Man1}: the energy $\varepsilon$ and
the strength of the electrostatic field $F$ are measured in units
of $\left|W_{0}\right|$ and in characteristic atom units,
respectively:

\[
\varepsilon = \frac{{W}}{{\left| {W_{0}}  \right|}}\,,\quad \quad \quad F =
\frac{{eE\hbar} }{{\left( {2m\left| {W_{0}}  \right|^{3}} \right)^{1/2}}}.
\]

Of course, the particle spin does not show itself in an
electrostatic external field. At that our equation (16) takes the
form ~\cite {Kad}

\begin{equation}
\left( { - \varepsilon}  \right)^{1/2} - 1 = \left( { -
\varepsilon _{0}} \right)^{1/2} + \frac{{F^{1/3}}}{{\pi} }{\kern
1pt} \left[ {Ai^{'}\left( {Bi^{'} + iAi^{'}} \right) + \varepsilon
_{0} F^{ - 2/3}Ai\left( {Bi + iAi} \right)} \right],
\end{equation}
where the common argument of the Airy functions $Ai, Bi$ and of
their derivatives $A{i}',B{i}'$ is $ - \varepsilon _{0} F^{-
2/3}$, and the zero approximation gives $\varepsilon _{0} = - 1$.
Formula (17), as it is affirmed in ~\cite {Man, Man1}, gives in
such an external field the transcendental equation

\begin{equation}
1 + \frac{{1}}{{\pi }}{\kern 1pt} F^{1/3}\left[ {Ai^{'}\left(
{Bi^{'} + iAi^{'}} \right) + \varepsilon F^{ - 2/3}Ai\left( {Bi +
iAi} \right)} \right] = 0.
\end{equation}
and the complex energy is contained in arguments of the Airy
functions $ - \varepsilon F^{-2/3}$. However, note once again that
it is impossible to pass from (17) to (21) by means of identical
transformations. The Airy functions on the left handed side of
equation (21) are finite while the integral on the right hand side
of (17) diverges. It is easy to see, at what stage of
transformations the infinity ``can disappear''. For example, it is
possible to pass from the general expression (17) to (21) with
help of transformation (15) with real $W_{0} $. However, the
integral on the left hand side of (15) diverges explicitly for
complex $W$ with a negative imaginary part, but the right handed
side remains finite. The mentioned in the previous section excess
of the accuracy in the derivation of equation (17) from the
comparison of asymptotics (18) and (19) reveals here.

At the same time, it is possible to receive the transcendental Eq.
(21) by analytical continuation of the right handed side of (20)
over the argument of the Airy functions into the complex plane,
replacing $\varepsilon_{0}$ by complex $\varepsilon$. As the
calculations show, values of $\varepsilon$ determined from (21)
are not differ too much from the values that gives the explicit
formula (20). In a weak field ($F \ll 1$) both relations reproduce
the well-known asymptotic expression ~\cite {Dem.1}

\[
 - \varepsilon = 1 + \frac{F^{2}}{16} +
\frac{i}{4}F\exp\left(- \frac{4}{3F} \right),
\]
if to be limited to first non-vanishing terms of the Airy
functions expansions. In the strong field $F = 1$ (i.e. comparable
with intra-atomic one) it is easy to get from (20)

\[
- \varepsilon \approx 1,0411 + 0,0451 \cdot i
\]
and the numerical solution of (21) gives ~\cite {Man, Man1}

\[
- \varepsilon \approx 1,0442 + 0,0388 \cdot i .
\]

Besides, there is a critical value of the electric field $F_{cr}$
at which $Re (-\varepsilon)= Im(-\varepsilon)$, i.e. the energy
gap between the ``shifted and broadened'' level and the continuous
spectrum disappears. Obviously, that usual concept of a
quasistationary level from ~\cite {Lan} loses the meaning for such
intensive electric fields and requires an additional definition.
In accordance with (20), that critical value is $F_{cr} \approx
13,26$. The transcendental equation (21) gives $F_{cr} \approx
11,38$.

Finelly note that there is a well-known analogy between the
complex energy in a theory of quasistationary system decays and
the complex dielectric permeability in physics of semiconductors
(for example, ~\cite {Ser, Asp, Rot, Nus}). Equations for the
complex dielectric permeability are derived, as a rule, on the
basis of well developed dispersive methods.

One more configuration of an external field which we consider is
essentially important to show the role of the magnetic field. For
simplicity we put the angle $\varphi = 0$ and average over
polarizations of bound particles. It is convenient to measure the
magnetic field like the electric one in natural atom units (see,
for example, ~\cite {Pop}):

\[
\quad h = \frac{eH\hbar}{mc\left|W_{0}\right|}.
\]

If $h \le F \ll 1$ it is possible to receive ~\cite {RoKr2} from
our equation (16)

\[
(- \varepsilon)^{1/2} - 1 = \frac{F^{2}}{32} + \frac{h^{2}}{48a} +
i\frac{F}{8} \left(1 + \frac{h}{3a F^{2}}\right) \exp \left(-
\frac{4b^{3/2}}{3F} \right).
\]

Here the dependent on spin factors $a$ and $b$ take values $a = -
2$, $b = 1 + h/2$ for a scalar particle and values $a = b = 1$ for
a spinor one. Therefore, the corrections to the complex energy due
to a magnetic field essentially depend on spins of particles. The
magnetic field decreases the width of a scalar particle bound
level and by that stabilizes it. On the contrary, the width of the
level of a particle with spin 1/2 increases in a magnetic field,
i.e. the latter strengthens the destabilizing action of an
electric field. That difference is still more noticeable in a
strong magnetic field ~\cite {RoKr1}. This field suppresses
exponentially the withdrawal of scalar particles from the $\delta
$-well, but intensifies it linearly for particles with spin:

\[
Im (- \varepsilon)^{1/2}_{\sigma = 0} \sim h^{1/2}\exp\left( -
\frac{2^{1/2} h^{3/2}}{3F} \right) \,;\quad \quad Im (-
\varepsilon)^{1/2}_{\sigma = 1/2} \sim \frac{h}{F^{1/3}}
Ai^{2}(F^{- 3/2}).
\]

If we distinguish particle polarizations after ionization it is
not difficult to see with help of equation (16) and condition (12)
that electrons with spins directed along a magnetic field
($a_{1}^{2} = 1/2, a_{2}^{2} = 0$) in the initially non-polarized
beam behave like scalar particles. Thus, the increasing of the
level width in a magnetic field is provided only by electrons with
spins oriented against the field.

The reason for the phenomenon described above can be seen from
equation (6) for the total energy of an electron in an external
field. Energies of Landau levels enhance with the increase of the
magnetic field. That complicates tunneling of a particle from the
bound state. Only one exception is the ground state of an electron
with spin directed against the magnetic field. The energy of this
state does not depend on \textit{H} at all, but its contribution
to the total probability of ionization increases with the increase
of the magnetic field.

Therefore for an adequate description of the influence of a
magnetic field on the behavior of fermionic quasistationary
systems it is necessary to take into account spin states of
fermions even if the considered problem is not relativistic.

\end{document}